
\magnification1200
\baselineskip=24pt
\belowdisplayskip=24pt
{\nopagenumbers
\topglue 1in
\centerline{NUMERICAL INVESTIGATION OF COSMOLOGICAL
SINGULARITIES}
\vskip 1in
\centerline{Beverly K. Berger$^{a,b}$ and Vincent Moncrief$^{a,c}$}
\vskip 1in
\centerline{$^a$Institute for Theoretical Physics, University of California,
Santa Barbara, CA 93106}
\centerline{$^b$Physics Department, Oakland University, Rochester, MI
48309}
\centerline{$^c$Department of Physics, Yale University, New Haven, CT
06511}
\vskip 1in
\centerline{July 21, 1993}
\vskip 1in
{\baselineskip=12pt PACS Nos:  04.20.Jb, 04.30.+x, 98.80.Dr

email:  berger@vela.acs.oakland.edu; moncrief@yalph2.bitnet}
\vfill
\eject

{\parindent=0pt ABSTRACT}

Although cosmological solutions to Einstein's equations are known to be
generically singular, little is known about the nature of singularities in
typical spacetimes.  It is shown here how the operator splitting used in a
particular symplectic numerical integration scheme fits naturally into the
Einstein equations for a large class of cosmological models (whose
dynamical variables are harmonic maps) and thus allows study of their
approach to the singularity.  The numerical method also naturally singles
out the asymptotically velocity term dominated (AVTD) behavior known
to be characteristic of some of these models, conjectured to describe
others, and probably characteristic of a subclass of the rest.  The method is
first applied to the generic (unpolarized) Gowdy T$^3$ cosmology.  Exact
pseudo-unpolarized solutions are used as a code test and demonstrate that
a 4th order accurate implementation of the numerical method yields
acceptable agreement.  For generic initial data, support for the conjecture
that the singularity is AVTD with geodesic velocity (in the harmonic map
target space) $< 1$ is found.  A new phenomenon of the development of
small scale spatial structure is also observed.  Finally, it is shown that the
numerical method straightforwardly generalizes to an arbitrary
cosmological spacetime on $T^3 \times R$ with one spacelike $U(1)$
symmetry.
\vfill
\eject}

{\parindent=0pt I.  Introduction.}

Powerful theorems [1] prove singularities to be a generic feature of
Einstein's equations yet say nothing about the nature of these singularities.
In particular, little is known about the singularity behavior of generic
spatially inhomogeneous cosmologies.  Belinskii, Khalatnikov, and
Lifshitz (BKL) [2] and coworkers [3] have long argued that the Mixmaster
dynamics [2, 4] of spatially homogeneous Bianchi Type VIII and IX
cosmologies [5] characterizes the generic ``big bang.''  Their results are
not generally accepted, however [6], and evidence suggests that
Mixmaster behavior disappears in models with more than three dynamical
degrees of freedom [7].  An alternative to Mixmaster dynamics is
asymptotically velocity term dominated (AVTD) behavior where
(heuristically) terms in Einstein's equations containing spatial derivatives
can be neglected in favor of those with time derivatives [8, 9].  Near the
singularity, AVTD solutions can be interpreted as a different spatially
homogeneous cosmology at each point in space.  The polarized Gowdy
cosmologies [10]  have been shown rigorously to belong to this class [9,
11,
12].  It has recently been conjectured that the general (unpolarized) Gowdy
models are AVTD [13].

We propose to study the approach to the singularity numerically
using a method uniquely suited to the task.  For both the Gowdy
cosmologies defined to have a symmetry plane (i.e.~two spatial,
hypersurface-orthogonal, surface-forming Killing fields) and the more
general cosmology possessing a single, spatial $U(1)$ symmetry [14], (at
least some of the) degrees of freedom can be understood as harmonic maps
[15].  The superhamiltonian whose variation yields these equations is just
an energy-like expression of the harmonic map fields [16].  The variation
of the ``kinetic'' term alone yields the AVTD equations of motion for these
fields (with all spatial-derivative-containing terms obtained upon variation
of the ``potential'' term).  This suggests that a symplectic scheme for
numerical integration that separately evolves the kinetic and potential
energy operators to approximate the total hamiltonian evolution is ideally
suited to this problem [17].  In the following discussion, we shall
demonstrate that this is indeed the case.

The approach to the singularity of the Gowdy T$^3$ cosmology
has been used to test the feasibility of our approach.  In the process of
code development and testing, we have been able to demonstrate AVTD
behavior of the singularity approach for several sets of (presumably)
generic initial data.  We have also noted that the solutions develop a
characteristic small scale spatial structure which represents a competition
between nonlinear growth and the approach to the AVTD regime which
freezes the spatial profile of the wave amplitudes.  This richness of the
Gowdy T$^3$ phenomenology will be discussed elsewhere [18].  Here we
wish to emphasize the applicability of our methods to the study of
cosmological singularities.

While the numerical study of plane symmetric cosmologies has
been underway since the late 1970's [19],  previous work tended to focus
on analogies between the cosmological problem and the original numerical
studies of colliding black holes [20].  This led to concentration on constant
mean curvature foliations and on the choice of lapse and shift.  Physical
interest centered on interacting wavepackets of a single polarization.  In
contrast, we begin with the predefined coordinate system in which the
equations are known to be relatively simple and within which the
harmonic map structure can be seen.  This foliation naturally selects (in
the plane symmetric case) a measure of area in the symmetry plane as the
time variable.  Our approach then easily allows study of unpolarized
Gowdy T$^3$ models and appears to be generalizable to the $U(1)$
problem.

In Section II, we shall describe the numerical method including its
generalization to arbitrary order of accuracy in both time [21] and space.
To date, our primary application of this method has been to the approach
to the singularity of the unpolarized Gowdy T$^3$ cosmology [10, 12, 16,
22].  In Section III, the relevant properties of this model will be reviewed.
Section IV demonstrates the validity of the code with a
``pseudo-unpolarized''
test model constructed by boosting in the harmonic map
target space an analytic solution for the polarized model.  In Section V, we
discuss results for a generic unpolarized model, demonstrating AVTD
behavior and briefly discussing the appearance of small scale spatial
structure.  In Section VI, the applicability of this method to the $U(1)$
problem is shown.  A summary and conclusions are given in Section VII.
\vfill
\eject
{\parindent=0pt II.  The Symplectic Integrator (SI) [17]}

For convenience, we shall restrict our discussion of the method to a
single degree of freedom which depends on only one spatial dimension
and time---$q(x,t)$ and its conjugate momentum $\pi(x, t)$.  We assume
that the equations of motion can be obtained by variation of a hamiltonian
$$H=\oint {dx}\left[ {{\textstyle{1 \over 2}}\pi ^2+V(q)}
\right].\eqno(2.1)$$

{\parindent=0pt Consider a differenced form of (2.1)}
$$H=\sum\limits_{i=0}^N {\left[ {{\textstyle{1 \over 2}}\left( {\pi _i^j}
\right)^2+V_i(q_k^j)} \right]}\eqno(2.2)$$

{\parindent=0pt where we assume periodic boundary conditions on the
lattice with labels $(i, j)$ denoting the point  $(x_i,t^j)$.  The potential
$V_i $ at the point  $x_i $ may depend on the value of $q$ at several
spatial grid points.}

The symplectic scheme splits the hamiltonian operator as
$$H=H_1+H_2\eqno(2.3)$$

{\parindent=0pt where}
$$H_1=\oint {dx}\,{\textstyle{1 \over 2}}\pi ^2\eqno(2.4a)$$

{\parindent=0pt and}
$$H_2=\oint {dx}\,V(q)\eqno(2.4b)$$

{\parindent=0pt respectively.  It is convenient to represent the scheme
using quantum mechanical notation.  It is based on the second order (in the
time-step $\varepsilon$) accurate approximant to the evolution operator:}
$$e^{-i\varepsilon H}=e^{-{(i/2)}
\varepsilon H_2}e^{-
i\varepsilon H_1}e^{-{(i/2)}
\varepsilon
H_2}+{\cal O}(\varepsilon ^3)\eqno(2.5)$$

{\parindent=0pt i.e.~to evolve $(\pi _i^j,q_i^j)$ at $(x_i,t^j)$ to $(\pi
_i^{j+1},q_i^{j+1})$ at $(x_i,t^j+\varepsilon )$, evolve with $H_2$ for
$1/2$ time-step, with $H_1$ for a full time-step, and with $H_2$ for
$1/2$
time-step using the appropriate intermediate result at each stage.  In this
evolution, $H_1$ and $H_2$ are separately to be regarded as the
hamiltonian of the system.  In the case where $H_1$ and $H_2$ can be
separately exactly solved, the implementation of the method becomes
trivial.}

For the hamiltonian (2.1), the scheme becomes [17]

$$\eqalign{&q_i^{j+1}=q_i^j+\varepsilon \,\left[ {\pi _i^j+{\textstyle{1
\over 2}}\varepsilon \,F_i(q_k^j)} \right]\cr
&\cr
  &\pi _i^{j+1}=\pi _i^j+{\textstyle{1 \over 2}}\varepsilon
\,F_i(q_k^j)+{\textstyle{1 \over 2}}\varepsilon
\,F_i(q_k^{j+1})\cr}\eqno(2.6)$$

{\parindent=0pt where $F_i(q_k^j)=-{{\partial V} \mathord{\left/
{\vphantom {{\partial
V} {\partial q_i^j}}} \right. \kern-\nulldelimiterspace} {\partial q_i^j}}$
is the appropriate force component.  As an example, we consider the wave
equation with
$$H_2=\sum\limits_{i=0}^N {{1 \over {2\Delta ^2}}\left( {q_{i+1}^j-
q_i^j} \right)^2}\eqno(2.7)$$

{\parindent=0pt where $\Delta$ is the lattice spacing in the $x$ direction.
Direct substitution shows that, in this case, the method is equivalent to the
standard leap-frog differenced form of the wave equation [23]}
$$q_i^{j+1}+q_i^{j-1}-q_{i+1}^j-q_{i-1}^j=0.\eqno(2.8)$$

{\parindent=0pt However, the SI algorithm has significant advantages
over the leap-frog scheme for our problem of the approach to the
singularity.}

{\parindent=0pt 1.  As a symplectic scheme, the evolution takes the form
of
a canonical transformation from the beginning to the end of the time-step
[17].  This may help to preserve the constraints of the cosmological
problem during the evolution.  (Although the continuum Einstein
equations automatically preserve the constraints during evolution, there is
no corresponding statement for the discretized equations.  Ultimately, this
is a consequence of the role of the constraints as the generators of
diffeomorphism invariance---a fundamental property of the continuum.)}

{\parindent=0pt 2.  In the cosmological case, $H_1$ is the hamiltonian
whose variation yields the AVTD equations.  If the solution is in the
AVTD regime, then this SI will become increasingly more accurate.}

{\parindent=0pt 3.  To avoid the problematic need to solve and resolve the
constraint equations at frequent intervals [24, 20], one can try to find an
accurate solution to the dynamical equations.  Suzuki has shown how to
generalize the SI to an arbitrary order in time [21].  The idea is to find an
approximant like (2.5) accurate to the desired order.  Such a program does
not have a unique solution.  Since it can be shown [21] that a $2m-1$
order accurate scheme is also $2m$ order accurate, one finds the
recurrence relation [21]}
$$S_{2m-1}(\varepsilon )=S_{2m}(\varepsilon )=S_{2m-
3}(k_m\varepsilon )S_{2m-3}[(1-2k_m)\varepsilon ]S_{2m-
3}(k_m\varepsilon )\eqno(2.9)$$

where
$$k_m=\left( {2-2^{{1 \mathord{\left/ {\vphantom {1 {(2m-1)}}} \right.
\kern-\nulldelimiterspace} {(2m-1)}}}} \right)^{-1}\eqno(2.10)$$

with
$$S_1(\varepsilon )=e^{-{(i/2)}\varepsilon H_2}e^{-
i\varepsilon H_1}e^{-{(i/2)}\varepsilon
H_2}.\eqno(2.11)$$

Thus the higher order scheme can be constructed from time-steps of the
appropriate duration of the second order scheme.}

The generalization of the spatial evolution to arbitrary order is
simple only for one spatial dimension.  For higher dimensions, the
construction must proceed on a case by case basis.  In the one spatial
dimension wave equation, the second derivative with respect to $x$ can be
obtained in a differenced form (for spatial grid interval $\Delta$) as

$${{d^2f} \over {dq^2}}=a_nf(q_i)+{1 \over {\Delta
^2}}\sum\limits_{k=1}^{n-1} {a_k[f(q_i+k\Delta )+f(q_i-k\Delta
)]}+{\cal O}(\Delta ^{2n}).\eqno(2.12)$$

{\parindent=0pt The coefficients are chosen to cancel the terms in the
Taylor expansion containing higher derivatives to the indicated order.  The
expression on the right hand side of (2.12) can be obtained as the negative
of the variation with respect to $f_i$  (where $f_i\equiv f(q_i)$) of

$$V=\sum\limits_{i=0}^N {\sum\limits_{k=1}^{n-1} {{{a_k} \over
{2\Delta ^2}}\left( {f_{i+k}-f_i} \right)^2}}\eqno(2.13)$$

where the $a_k$ are the same as those in (2.12) and with $a_n=-
2\sum\nolimits_{k=1}^{n-1} {a_k}$.  Thus the second order accurate
expression is (2.7) with the 4th order accurate one given by

$$V=\sum\limits_{i=0}^N {\left[ {{2 \over {3\Delta ^2}}\left( {q_{i+1}-
q_i} \right)^2-{1 \over {24\Delta ^2}}\left( {q_{i+2}-q_i} \right)^2}
\right]},\eqno(2.14)$$

etc.  This prescription is easily extended to more general potentials;  e.g.
$$V=\oint {dx{1 \over 2}F(q)\,q^2}\eqno(2.15)$$

is differenced as
$$V=\sum\limits_{i=0}^N {\sum\limits_{k=1}^{n-1} {{{a_k} \over
2}F[{\textstyle{1 \over 2}}(q_{i+k}+q_i)]\left( {q_{i+k}-q_i}
\right)^2}}.\eqno(2.16)$$

For two spatial dimensions, generalization to the Laplacian is trivial.
However, potentials with cross-terms or different coefficients for
${{\partial ^2} \mathord{\left/ {\vphantom {{\partial ^2} {\partial x^2}}}
\right. \kern-\nulldelimiterspace} {\partial x^2}}$ and ${{\partial ^2}
\mathord{\left/ {\vphantom {{\partial ^2} {\partial y^2}}} \right. \kern-
\nulldelimiterspace} {\partial y^2}}$ cannot be differenced to higher order
by this prescription due to the difficulty of eliminating higher derivative
terms in a multidimensional Taylor expansion.}
\vfill
\eject
{\parindent=0pt III.  The Gowdy T$^3$ Universe Test Case.}

The Gowdy T$^3$ cosmology is conveniently described by the
metric [16]
$$ds^2=e^{{\lambda  \mathord{\left/ {\vphantom {\lambda  2}} \right.
\kern-\nulldelimiterspace} 2}}e^{{\tau  \mathord{\left/ {\vphantom {\tau
2}} \right. \kern-\nulldelimiterspace} 2}}(-\,e^{-2\tau }\,d\tau ^2+d\theta
^2)+e^{-\tau }\,[e^Pd\sigma ^2+2e^PQ\,d\sigma \,d\delta +(e^PQ^2+e^{-
P})\,d\delta ^2]\eqno(3.1)$$

{\parindent=0pt where $\lambda$, $P$, and $Q$ are functions of $\theta$
and $\tau$ only.  The T$^3$ spatial topology is imposed by requiring the
angular coordinates $\sigma$ and $\delta$ to have arbitrary finite range
and $0 \le \theta \le 2\pi$.  The time variable $\tau$ measures the area in
the symmetry plane and $\to \infty$ at the singularity [9, 12].  The
physical interpretation of the polarized model ($Q \equiv 0$) has been
discussed extensively [12, 25].  The independent Einstein equations from
(3.1) are

$$P,_{\tau \tau }-\;e^{-2\tau }P,_{\theta \theta }-\;e^{2P}\left( {Q,_\tau
^2-\;e^{-2\tau }Q,_\theta ^2} \right)=0,\eqno(3.2)$$

$$Q,_{\tau \tau }-\;e^{-2\tau }Q,_{\theta \theta }+\;2\left( {P,_\tau Q,_\tau
^{}-\;e^{-2\tau }P,_\theta Q,_\theta ^{}} \right)=0,\eqno(3.3)$$

$$\lambda ,_\theta -\;2(P,_\theta P,_\tau +\;e^{2P}Q,_\theta Q,_\tau
)=0,\eqno(3.4)$$

and
$$\lambda ,_\tau -\;[P,_\tau ^2+\;e^{-2\tau }P,_\theta ^2+\;e^{2P}(Q,_\tau
^2+\;e^{-2\tau }\,Q,_\theta ^2)]=0.\eqno(3.5)$$

The latter two equations for the background $\lambda(\theta,\tau)$ are
respectively the $\theta$-momentum and hamiltonian constraints.  [The
T$^3$ topology requires the integral of (3.4) to vanish---equivalent to
requiring zero total $\theta$-momentum.]}

Since the ``wave'' equations (3.2) and (3.3) do not contain
$\lambda$, their evolution is unconstrained.  It has been shown, albeit for
a different set of variables defined by [16]
$$\eqalign{e^P=&\cosh W+\sinh W\cos \Phi \cr
  e^PQ=&\sinh W\sin \Phi \cr}\eqno(3.6)$$

{\parindent=0pt that the ``wave'' equations are harmonic map equations for
the metric
$$dS^2=dP^2+\;e^{2P}dQ^2.\eqno(3.7)$$

This is just the harmonic map property of the fields $P$ and $Q$ with
(3.7) the metric of the target space [15].  The field equations can then be
derived from the hamiltonian

$$H={1 \over 2}\oint {d\theta \left[ {\pi _P^2+\;e^{-2P}\pi _Q^2+\;e^{-
2\tau }\left( {P,_\theta ^2+\;e^{2P}Q,_\theta ^2} \right)}
\right]}.\eqno(3.8)$$

Clearly, (3.8) is in the form required by the SI algorithm.}

We note here for future reference that the metric (3.7) admits three
Killing fields corresponding to the transformations (for constant parameter
$\rho$)

$$\left\{ \matrix{P\to P\hfill\cr
  Q\to Q+\rho \hfill\cr} \right.\eqno(3.9)$$

$$\left\{ \matrix{P\to P-\ln \rho \hfill\cr
  Q\to \rho \,Q\hfill\cr} \right.\eqno(3.10)$$

{\parindent=0pt and
$$\left\{ \matrix{e^P\to {\textstyle{1 \over 2}}\left[ {e^P(1+Q^2)+\;e^{-
P}} \right]+{\textstyle{1 \over 2}}\left[ {e^P(1-Q^2)-\;e^{-P}} \right]\cos
\rho -\;e^PQ\sin \rho \hfill\cr\hfill\cr
  e^PQ\to e^PQ\cos \rho +{\textstyle{1 \over 2}}\left[ {e^P(1-Q^2)-\;e^{-
P}} \right]\sin \rho \hfill\cr} \right.\eqno(3.11)$$

This last apparently complicated transformation is just $W\to W,\quad\Phi
\to \Phi +\Phi _0$ in the other coordinates.  The presence of the factor
$e^{-2\tau }$ in (3.8) and in the wave equations suggests that as the
singularity at $\tau = \infty$ is approached, the spatial derivatives can be
neglected yielding the AVTD solution.  In the absence of the spatial
derivative terms, Eqs. (3.2) and (3.3) can be solved exactly in terms of
four arbitrary constants $\alpha$, $\beta$, $\zeta$, and $\xi$ as
$$P=\ln [\alpha \,e^{-\beta \tau }(1+\zeta ^2e^{2\beta \tau
})]\eqno(3.12)$$

and
$$Q=-\;{{\zeta \,e^{2\beta \tau }} \over {\alpha \,(1+\zeta ^2e^{2\beta \tau
})}}+\xi .\eqno(3.13)$$

Substitution of (3.12) and (3.13) into the AVTD form of (3.4) and (3.5)
yields
$$\lambda =-\beta ^2\tau +\lambda _0.\eqno(3.14)$$

As $\tau \to \infty$, (3.12) and (3.13) become
$$P=\beta \tau \quad;\quad Q=Q_0\eqno(3.15)$$

with $Q_0={1 \mathord{\left/ {\vphantom {1 {\alpha \zeta +\xi }}} \right.
\kern-\nulldelimiterspace} {\alpha \zeta +\xi }}$.  If a Gowdy solution
approaches the AVTD limit, one expects it to have the form (3.12)--(3.14)
with (in general) different values of $\alpha$, $\beta$, $\zeta$, and $\xi$ at
each value of $\theta$.  For the polarized case ($Q = 0$), it was shown
[12] that substitution of (3.15) as the limiting form of the exact solution in
the metric (3.1) yields the Kasner solution [26] with $\theta$ dependent
Kasner parameter.  In the general case ($Q\ne 0$), the Kasner axes are
rotated with respect to the coordinate axes.  Isenberg and Moncrief have
shown [9] in the polarized case that every solution is AVTD.  It is
conjectured [13] that this is also true in the unpolarized model.}

In the following discussion, we shall consider only the wave
equations (3.2) and (3.3) since the background $\lambda(\theta,\tau)$ may
be easily constructed after the dynamical $P$ and $Q$ have been found.
That $P$ and $Q$ are amplitudes for the two orthogonal polarizations of
gravitational waves may be seen by analogy with linearized gravity [12,
27].  If the metric $g_{\mu \nu }$ is expressed as
$$g_{\mu \nu }=\gamma _{\mu \nu }^{(0)}+h_{\mu \nu }^{(1)}+k_{\mu
\nu }^{(2)}\eqno(3.16)$$

{\parindent=0pt with $P$ and $Q$ assumed small then
$$\gamma _{\mu \nu }^{(0)}={\hbox{\rm diag}} (-e^{{\lambda
\mathord{\left/
{\vphantom {\lambda  2}} \right. \kern-\nulldelimiterspace} 2}}e^{{{-
3\tau } \mathord{\left/ {\vphantom {{-3\tau } 2}} \right. \kern-
\nulldelimiterspace} 2}},\,e^{{\lambda  \mathord{\left/ {\vphantom
{\lambda  2}} \right. \kern-\nulldelimiterspace} 2}}e^{{\tau
\mathord{\left/ {\vphantom {\tau  2}} \right. \kern-\nulldelimiterspace}
2}},\,e^{-\tau },e^{-\tau })\eqno(3.17a)$$

describes a background metric.  The designation as background can be
enhanced by the introduction of spatial averaging [27, 28, 25].  We also
find that
$$h_{\mu \nu }^{(1)}=e^{-\tau }\,P\,\varepsilon _{+\mu \nu }+e^{-\tau
}\,Q\,\varepsilon _{\times \mu \nu }\eqno(3.17b)$$

where $\varepsilon _+\;{\hbox {\rm and}}\;\varepsilon _\times $ are the
gravitational wave polarization tensors.  In the $\sigma$-$\delta$ plane,
$$k_{\mu \nu }^{(2)}=e^{-\tau }\left( {\matrix{{{\textstyle{1 \over
2}}P^2}&{PQ}\cr
{PQ}&{Q^2+{\textstyle{1 \over 2}}P^2}\cr
}} \right)\eqno(3.17c)$$

with all other components zero.  It is easy to see [12, 27] that in zeroth
order, the waves act as sources for the background in Eqs. (3.4) and (3.5).
In first order, $P$ and $Q$ satisfy linear wave equations (3.2) and (3.3).
The nonlinearities of the waves enter at the next order.  In some sense, we
cannot consider these terms to be higher order since the solution is
qualitatively different if both polarizations are present even if the
amplitude of $Q$ is small [29].  This is explained by the scaling symmetry
(3.10) which implies that a solution to (3.2) and (3.3) is independent of the
amplitude of $Q$ as long as it is non-zero.}

The polarized case (e.g. $Q = 0$) yields a single linear wave
equation with the general exact solution [12]
$$P(\theta ,\tau )=\sum\limits_{n=0}^\infty  {Z_0(ne^{-\tau })(a_n\cos
n\theta +b_n\sin n\theta )}\eqno(3.18)$$

{\parindent=0pt where $Z_0(x)$ is a general solution to Bessel's equation
of zero order and the $a$'s and $b$'s are arbitrary constants. It is possible
to use this exact solution to generate exact ``pseudo-unpolarized''
solutions to the unpolarized equations (3.2) and (3.3) by
means of the boost symmetry (3.11).  Given an exact solution $P_0$ from
(3.18), we obtain a class of solutions}

$$\eqalign{e^P=&\cosh P_0+\sinh P_0\cos \rho \cr
  e^PQ=&\sinh P_0\sin \rho \cr}\eqno(3.19)$$

{\parindent=0pt for all values of the parameter  $0 \le \rho \le 2\pi$.
Perhaps the simplest of these solutions is

$$\eqalign{&P=\ln \cosh P_0\cr
  &Q=\tanh P_0\cr}\eqno(3.20)$$

found for $\rho = {\pi/2}$.  The pseudo-unpolarized solutions make
excellent code tests since the fact that they are non-generic is not apparent
to the computer.  Direct substitution of (3.20) into the equations of
motion (3.2) and (3.3) shows that nonlinear terms from variation of both
$H_1$ and $H_2$ [see the discussion in Section II and Eq.~(3.8 )] must
cancel corresponding expressions which arise in the linear terms.  Thus the
entire code is tested.}

Grubi\v si\'c and Moncrief [13] have defined several functions of
$P$ and $Q$ and their conjugate momenta which become constant in
$\tau$ in the AVTD regime.  They also predict the rate of decay to the
AVTD regime in terms of the $\theta$ dependent parameters $\alpha$,
$\beta$, $\zeta$, and $\xi$.  For the purposes of this paper, we shall
consider only the parameter

$$v\equiv \left( {P,_\tau ^2+\;e^{2P}Q,_\tau ^2} \right)^{1\/
2}\eqno(3.21)$$

{\parindent=0pt which represents geodesic velocity in the target space
(3.7).  Grubi\v si\'c and Moncrief have conjectured that for a generic
Gowdy T$^3$ model, the
AVTD regime will be characterized by $0 \le v < 1$ for all $\theta$.  The
restriction to generic models must be made because any value of $v$ is
allowed for polarized solutions [9] and $v = 1$ is achieved for some
non-polarized
but ``asymptotically polarized'' solutions [30].  (We note
that $v$ is invariant under the transformations (3.9)--(3.11) so that
pseudo-unpolarized
solutions can also have any value of $v$.)  The heuristic basis
for this conjecture is easily seen.  The term $e^{-2\tau }e^{2P}Q,_\theta
^2$ in (3.2) becomes $e^{-2\tau (1-v)}Q,_\theta ^2$ in the AVTD limit.
Clearly, if $v > 1$ (with $Q,_\theta$ fixed), this term will grow, contrary
to the AVTD assumption that it is negligible.}
\vfill
\eject
{\parindent=0pt IV.  The Pseudo-Unpolarized Test Case.}

In differenced form, the hamiltonian (3.8) becomes

$$\eqalign{H=&{1\over 2}\sum\limits_{i=1}^N {\left( {\pi _{P_i}^2+\,
e^{-2P_i}\pi_{Q_i}^2} \right)}\cr
  &+{{e^{-2\tau }} \over {(\Delta \theta )^2}}\sum\limits_{i=0}^N
{\left\{ {a\left[ {\left( {P_i-P_{i-1}} \right)^{2.}+e^{P_i+P_{i-1}}\left(
{Q_i-Q_{i-1}} \right)^2} \right]} \right.}\cr
  &+\,\left. {b\,\left[ {\left( {P_i-P_{i-2}} \right)^{2.}+e^{P_i+P_{i-
2}}\left( {Q_i-Q_{i-2}} \right)^2} \right]} \right\}\cr}\eqno(4.1)$$

{\parindent=0pt where $(a, b) = (1/2, 0)$ and $(2/3, - 1/24)$ yields
equations correct to second and fourth order respectively in the spatial
derivatives.  [Extension to sixth order requires corresponding terms
$(P_{i+3}-P_i)^2$, etc.~with coefficients $(a, b, c) = (3/4, - 3/40,
1/180)$.]  The first sum is $H_1$ and the second $H_2$.  To evolve with
$H_1$, solve the AVTD solution [(3.12), (3.13), and their
$\tau$Jderivatives] for a given $P$ and $Q$ and their conjugate momenta
(at each $\theta$ value) for the parameters $\alpha$, $\beta$, $\zeta$, and
$\xi$.  Use these parameters to propagate the initial data to the end of the
time-step.  The evolution with $H_2$ is even easier since it contains no
momenta so that $P_i$ and $Q_i$ remain constant.  The momenta are
evolved with the (now constant) gradients of $H_2$.  The overall time
dependent factor $e^{-2\tau }$ is then trivially integrated.  (We note that
one may alternatively treat $\tau$J as an extra degree of freedom.)
Suzuki's method [21] is used to ensure that the time evolution is accurate
to the desired order.  Greater details of our algorithm are given in the
Appendix.}

As a code test, initial data appropriate to the pseudo-unpolarized
boost (3.20) of the exact solution to (3.18) given by
$$P_0(\theta ,\tau )=Y_0(e^{-\tau })\cos \theta \eqno(4.2)$$

{\parindent=0pt where $Y_0(x)$ is an irregular Bessel function were
evolved numerically toward the singularity [31].  The exact boosted
solution is displayed in Fig.~1.  Note that the boost transformation has
generated large $\theta$ derivatives (particularly in $Q$) due to the
hyperbolic tangent.  Figures 2 and 3 illustrate the differences between the
numerical and exact solutions for the 2nd and 4th order schemes
respectively.  Although the errors in the second order algorithm are small
almost everywhere ( $\approx 1\%$), they become unacceptably large
($\approx 1$) as $\tau$ increases in the regions of large $\theta$
derivative.  Improvement with the 4th order scheme is dramatic with
relative errors $\approx 10^{-5}$ everywhere.}

The accuracy of the 4th order code appears to be acceptable for the
following reasons:  The code tests were run with low spatial resolution of
100 total grid points with 99 (97) representing $[0, 2\pi]$ for 2nd (4th)
order (due to the imposition of periodic boundary conditions).  The range
of $\tau$ was between 0 and $\approx 23$.  The $\tau$ interval was
chosen for convenience---there was no barrier to a much closer approach
to the singularity.  The large spatial gradients (in $Q$) were correctly
represented by the code.  We shall see (next section) that such features are
in fact also characteristic of the generic Gowdy T$^3$ solution.
\vfill
\eject
{\parindent=0pt V.  Results for a ``Generic'' Unpolarized Gowdy T$^3$
Model.}

Here we shall discuss the results from a single initial Gowdy T$^3$
data set for evolution toward the singularity.  Comparison to other sets
appears to indicate that the behavior we report here is typical for standing
wave initial data.  Traveling waves will be discussed elsewhere as will the
full range of Gowdy T$^3$ phenomenology [18].  Following a suggestion
by Chrusciel [32], we consider initial data for which the parameter $v$ in
(3.20) exceeds unity with the initial time dependence that of the AVTD
behavior (3.15).  The conjecture [13] is that a true AVTD regime with $v
< 1$ everywhere will arise during the course of the evolution.  The
selected initial data were
$$P=0,\;\;\pi _P=10\cos \theta ,\;\;Q=\cos \theta ,\;\;\pi _Q=0\eqno(5.1)$$

{\parindent=0pt so that $v_0=10\;\left| {\cos \theta } \right|$.  The
evolution was performed with 800 spatial grid points for the range $0 \le
\tau \le 6\pi$ with the 4th order accurate code.  To study the approach to
the AVTD regime, we average reported (i.e. saved rather than computed)
values of $P$, $Q$, and $v$ over nearest and next nearest neighbors (to
avoid grid scale size structures which must be regarded to be unreliable).
Figures 4--6 illustrate $P$, $Q$, and $v$ respectively vs.~$\theta$ for
various values of $\tau$.  The approach to the limiting AVTD behavior
(3.15) is seen clearly in Fig.~7 which shows $P$, $Q$, and $v$ vs. $\tau$
at selected values of $\theta$.  The entire evolution is displayed in a 3-D
surface plot in Figures 8 (for $P$ and $Q$) and 9 (for $v$ and $v < 1$).}

We note the following features of the evolution:

{\parindent=0pt 1.  The wave amplitude $P$ develops a complicated
spatial profile which then freezes after which $P$ grows linearly without
change of shape.}

{\parindent=0pt 2.  The amplitude $Q$ initially grows rapidly [where
$\cos \theta <0$ since there $P,_\tau $ in (3.3) acts as inverse damping]
but then becomes constant.  (There is some spatial structure in $Q$ which
does not show up for the amplitude scale used in this graph.)}

{\parindent=0pt 3.  The parameter $v$ decays essentially monotonically to
values $< 1$ everywhere.  This is emphasized by Fig.~9b which has a
scale adjusted to display only $v \le 1$.}

{\parindent=0pt It thus appears that the AVTD regime has been reached
and that the parameter $v$ has fallen below unity as conjectured.
Although we have begun with a single spatial mode, we expect the
nonlinear terms to cause a generic evolution.  Other standing wave initial
data have been examined.  It appears that the evolution is controlled to
some extent by $v_0$, the initial value of the parameter $v$.  If $v_0<1$,
there is little growth of spatial structure and the AVTD regime is reached
quickly.  The larger $v_0$, the more nonlinear the wave interactions will
be.}

If spatial averaging is not used, $v>1 $ can occur at isolated points
where $Q,_\theta \approx 0$.  (See Fig.~10.)  Such points, in effect,
represent the locally polarized models where $v \ge 1$ is allowed.  To the
extent that $Q,_\theta \ne 0$ (i.e. that the model really is generic at that
point), it still evolves to reach the AVTD regime as conjectured at some
$\tau >> 6\pi$.  The spatial averaging dilutes the influence of spiky
features in the dynamical variables.  The development of AVTD behavior
with spatial averaging removed is shown in Fig.~11 where $P/\tau$ and
$v$ over a limited range in $\theta$ are shown for $\tau = 6\pi$ and
$\tau = 14\pi$.  The curves (at each $\tau$ value) should be identical in
the AVTD limit as $\tau \to \infty$ [see (3.15)].  It appears that the
asymptotic AVTD behavior will be achieved at sufficiently large $\tau$.

The growth of spatial structure at
arbitrarily small scales appears to be characteristic of the Gowdy T$^3$
dynamics for $v_0\ge 1$.  Figure 12 reproduces Fig.~8a for $0 \le \tau \le
4.2$.  It is easy to see that the initial $cos \theta$ spatial profile
nonlinearly
generates [through (3.2)] $cos 2\theta$ dependence, etc.  The development
of this structure ends when the AVTD limit is reached.  The competition
between  nonlinear growth and the spatial freezing of AVTD behavior
suggests that there may exist a $v_0$ dependent time-scale to characterize
the phenomenon.  It is possible that this small scale structure
may be related to the critical
phenomena observed by Choptuik [33] and later by Abrahams and Evans
[34] for spherical collapse of a scalar field and axisymmetric collapse of
gravitational waves respectively.

The presence of this small scale spatial structure which eventually
reaches the grid spacing scale (unless the AVTD regime is reached) causes
the detailed numerical evolution to become dependent on the chosen
spatial resolution---i.e. at a given $\tau$, the finer the grid, the smaller
the
feature that is seen.  However, for any $\tau$ there exists a spatial
resolution which is sufficient to resolve all the small scale features.  This
is shown (with no spatial averaging) in Fig.~13 which compares the same
feature (in this case for $\pi_P$) at various spatial resolutions.  We note
that the feature is completely resolved at 6400 grid points with a profile
that agrees completely with that obtained for 1600 grid points.  Greater
deviations are found for coarser grids.  All resolutions represent the
solution where it is smooth.  The evolution of this same feature is shown
in Fig.~14.  We see that the feature has narrowed (and decreased in
amplitude) and is no longer resolved at 6400 grid points.  This narrowing
and decreasing amplitude explains the apparent decrease in spatial
structure seen in the evolution shown in Fig.~4 (although some is due to
the increased range in the amplitude of $P$).  The spatial averaging used
in Figs.~4--9 washes out the structure at small spatial scales.  Subsequent
evolution of the feature in Fig.~14 shows little change, indicating
that AVTD behavior (where $\pi_P \to constant$) is arising.

This small scale structure is almost certainly a real property of the
equations rather than a numerical artifact since it can be resolved with
sufficient spatial resolution.  (As a further code test, the structure is seen
to
disappear when the code is run backward in time.)  It was also seen in
studies of the approach to the singularity using a completely different
numerical
algorithm [35].)  Its characterization is currently under investigation.  Its
presence may signal a requirement for adaptive gridding [36] to achieve
the spatial resolution that appears to be necessary.
\vfill
\eject

{\parindent=0pt VI.  The $U(1)$ Problem.}

It has been shown [14] that an arbitrary cosmological spacetime on
$T^3\times R$ containing a spacelike $U(1)$ symmetry can be described
by the 5 degrees of freedom ${\varphi ,\omega ,x,z,\Lambda }$ and their
respective conjugate momenta ${p,r,p_x,p_z,p_\Lambda }$.  All variables
are functions of the spatial coordinates $u$ and $v$ and a time coordinate
$\tau$ which measures the size of the universe in the symmetry direction.
The conformal 2-metric of the space orthogonal to the Killing field is
$$e_{ab}={1 \over 2}\left( {\matrix{\matrix{e^{2z}+\;e^{-
2z}(1+x)^2\quad\hfill\cr
  \hfill\cr}&\matrix{e^{2z}+\;e^{-2z}(x^2-1)\hfill\cr
  \hfill\cr}\cr
{e^{2z}+\;e^{-2z}(x^2-1)\quad}&{e^{2z}+\;e^{-2z}(x-1)^2}\cr
}} \right)\eqno(6.1)$$

{\parindent=0pt with $a, b = 1, 2$ and $\det e_{ab}=1$.  (The 2-metric
itself is $g_{ab}=e^\Lambda\,e_{ab}$.)  The scalar
curvature of this conformal 2-space is
$$\eqalign{{}^{(2)}R(e_{ab})\equiv R=&[e^{-2z}(1-x)x,_u+e^{-2z}(1-
x)^2z,_u-e^{2z}z,_u-e^{-2z}x,_v\cr
  \cr
  &+e^{2z}z,_v+e^{-2z}z,_v+e^{-2z}xx,_v-e^{-2z}x^2z,_v],_u\cr
  \cr
  &+[-e^{-2z}(1+x)x,_v+e^{-2z}(1+x)^2z,_v-e^{2z}z,_v+e^{-2z}x,_u\cr
  \cr
  &+e^{2z}z,_u+e^{-2z}z,_u+e^{-2z}xx,_u-e^{-
2z}x^2z,_u],_v\quad.\cr}\eqno(6.2)$$

Given these definitions, the field equations (for the dynamical variables in
the gauge $N=\sqrt {{}^2g}=e^\Lambda $ where $N$ is the lapse and
$^2g$ is the determinant of the 2-metric) can be derived from the
hamiltonian
$$\eqalign{H=&-\;\oint \oint{du\,dv}\,\left( {{\textstyle{1 \over
8}}p_z^2+{\textstyle{1 \over 2}}e^{4z}p_x^2+{\textstyle{1 \over
8}}p_{}^2+{\textstyle{1 \over 2}}e^{4\varphi }r_{}^2-{\textstyle{1
\over 2}}p_\Lambda ^2} -2\,p_\Lambda\right) \cr
  \cr
 & \;-\;{e^{-2\tau}}\,\oint \oint {du\,dv}\left[ {-e^\Lambda R+e^\Lambda
 \left( {e^{ab}\Lambda ,_a} \right),_b+2e^\Lambda e^{ab}\varphi ,_a
\varphi
,_b+{\textstyle{1 \over 2}}e^\Lambda e^{-4\varphi }e^{ab}\omega
,_a\omega ,_b} \right]\quad.\cr}\eqno(6.3)$$

We note that the integrand in Eq.~(6.3)
is not the superhamiltonian, but differs from it by an overall sign
and an additional term linear in $p_\Lambda$.  The sign arises from the
fact that $\tau$
results from an original time variable $t$ via $dt=-\,e^{-\tau }d\tau $ while
the additional term comes from the fact that a time dependent
transformation is required to obtain our variables from the original
Arnowitt, Deser, and Misner (ADM) [37] ones.}

It is clear that the hamiltonian (6.3) fits naturally into the form of
the SI operator splitting.  Even more striking is the fact that the first
integral, $H_1$, contains two copies of the Gowdy T$^3$ kinetic term
[see
(3.8)] plus the kinetic energy of a free particle (with the ``wrong'' sign).
Thus we already have the exact solution for $H_1$ from (3.12) and (3.13)
as in the Gowdy case plus the trivial free particle solution.  The second
integral, $H_2$, is more difficult---not to solve since there are no
momenta
so all the variables are assumed constant over the sub-time step---but to
spatially difference to the correct order of accuracy.

In addition, the dynamical degrees of freedom are constrained---the
Gowdy split into wave and background variables does not occur.  For
example, the integrand of (6.3), leaving out the term linear
in $p_\Lambda$, is proportional to the hamiltonian
constraint.  There are also non-trivial momentum constraints.  Since the
same terms occur in $H_2$ and in the hamiltonian constraint, it is
probably
advantageous to preserve the divergence structure of the first two terms in
the square bracket rather than to partially integrate them.  A differencing
scheme for the field equations that recognizes the underlying structure of
the constraints may aid in keeping the numerical evolution  on the
constraint hypersurface.  Although the differencing scheme for $H_2$
outlined in Section II can be extended to the Laplacian, it becomes
problematical for the mixed partial derivatives that will arise here.  We
plan to begin with a plausible 2nd order accurate scheme.

Since the dynamical evolution is constrained, specification of
initial data is non-trivial.  Fortunately, examples of ``half-polarized'' exact
solutions of the constraint equations are known [38].

Although almost nothing is known about the generic behavior of
the $U(1)$ models, it is known that not all solutions can be AVTD.  This
is because this class of solutions contains the Mixmaster cosmologies [2--
4] which (since the influence of the potential never disappears) are not
AVTD.  If, as has been conjectured, the Mixmaster dynamics cannot
survive the presence of the spatial inhomogeneity degrees of freedom, the
models could still be AVTD.  We note here, however, that the
transformation necessary to obtain our variables has obscured the meaning
of AVTD since the transformation to the twist potential degree of freedom
$[\omega, r]$ has interchanged coordinates and momenta.  It is also
expected that small scale spatial structure will appear in the generic case.
The possible need for high spatial resolution in two spatial dimensions
may
push the limits of current computational technology.
\vfill
\eject
{\parindent=0pt VII.  Summary and Conclusions.}

A SI scheme that splits the hamiltonian into exactly solvable
kinetic and potential pieces is ideally suited to the numerical study of the
singularity structure of spatially inhomogeneous cosmologies.  For both
the Gowdy T$^3$ model and the more general $U(1)$ problem (on
$T^3\times R$), the dynamical equations arise from a variational principle
that also splits naturally into kinetic and potential pieces which separately
can be solved exactly.  The exact solution for the kinetic sub-hamiltonian
is in fact just the AVTD solution conjectured to arise for the generic
Gowdy T$^3$ model (taken twice plus a free particle term in the $U(1)$
problem).

The SI code has been implemented through 4th order in both time
(using Suzuki's method) and space for the generic Gowdy model.
Comparison for a pseudo-unpolarized test case (obtained by a boost
symmetry from an exact solution for the polarized model) shows
agreement to $1:10^5$ for the 4th order accurate code.  The 2nd order
accurate code diverges unacceptably in regions of large spatial derivative
as the singularity is approached.  The cost of the extra accuracy is
essentially a factor of three in computational time (since three 2nd order
time steps are required to produce a 4th order time step).  The extra spatial
accuracy involves negligible cost.

For reasonably generic Gowdy initial data, we have been able to
support the conjecture [13] that the models are AVTD with $v < 1$
everywhere.  More detailed study of the approach to the AVTD regime to
compare with the detailed predictions in [13] is in progress.

An interesting new phenomenon of the development of small scale
spatial structure has been observed.  Studies to characterize this behavior
in terms of the competition between nonlinear generation of short
wavelength modes and the freezing of the spatial profile in the AVTD
regime are underway.

Thus we have applied the SI scheme to the unpolarized Gowdy
T$^3$ cosmology and have been able to test the code, to study and verify
AVTD regime conjectures, and to discover the new phenomenon of
nonlinear small scale structure.  An even richer phenomenology awaits the
application of this method to the unknown territory of the U(1) problem.
\vskip 1in

{\parindent=0pt ACKNOWLEDGMENTS }

This research was supported in part by NSF Grants PHY89-04035 to the
Institute for Theoretical Physics, University of California / Santa Barbara,
PHY91-07162 to Oakland University, and PHY92-01196 to Yale
University.  Computations were performed using the facilities of the
National Center for Supercomputing Applications at the University of
Illinois. BKB wishes to thank the Institute for Geophysics and Planetary
Physics at Lawrence Livermore National Laboratory for hospitality.  The
authors wish to thank Boro Grubi\v si\'c, David Garfinkle, and Salman
Habib
for useful discussions.
\vfill
\eject
{\parindent=0pt Appendix:  Details of the Second Order Accurate SI
Algorithm for the Gowdy T$^3$ Model.}

The hamiltonian (3.8) can be put in differenced form

$$\eqalign{H=&\sum\limits_{i=0}^N {{\textstyle{1 \over 2}}\left( {\pi
_{P_i}^2+e^{-2P_i}\pi _{Q_i}^2} \right)}\cr
  &+\sum\limits_{i=0}^N {{\textstyle{1 \over 2}}e^{-2\tau }\left[ {\left(
{{{P_{i+1}-P_i} \over {\Delta \theta }}} \right)^2+e^{P_{i+1}+P_i}\left(
{{{Q_{i+1}-Q_i} \over {\Delta \theta }}} \right)^2}
\right]}\cr}\eqno(A.1)$$

{\parindent=0pt with each sum regarded to be an independent
sub-hamiltonian ($H_1$ and $H_2$ respectively).  Given initial data
${Q_i(\tau ^j),P_i(\tau ^j),\pi _{Q_i}(\tau ^j),\pi _{P_i}(\tau ^j)}$, we use
$H_2$ to evolve to $\tau ^j+{\raise3pt\hbox{$\scriptstyle 1$}
\!\mathord{\left/ {\vphantom {\scriptstyle {1 2}}}\right.\kern-
\nulldelimiterspace} \!\lower3pt\hbox{$\scriptstyle 2$}}\Delta \tau $ to
yield
$$\eqalign{&\tilde Q_i=Q_i(\tau ^j),\cr
&\cr
  &\tilde P_i=P_i(\tau ^j),\cr
&\cr
  &\tilde \pi _{Q_i}=\pi _{Q_i}(\tau ^j)+{\textstyle{1 \over 2}}e^{-2\tau
_j}\left[ {e^{-(\tau _{j+1}-\tau _j)}-1} \right]\left. {{{\partial V} \over
{\partial Q_i}}} \right|_{{Q_k(\tau ^j),P_k(\tau ^j)}},\cr
&\cr
  &\tilde \pi _{P_i}=\pi _{P_i}(\tau ^j)+{\textstyle{1 \over 2}}e^{-2\tau
_j}\left[ {e^{-(\tau _{j+1}-\tau _j)}-1} \right]\left. {{{\partial V} \over
{\partial P_i}}} \right|_{{Q_k(\tau ^j),P_k(\tau ^j)}}\cr}\eqno(A.2)$$

where $V$ is the ``spatially dependent'' part of the potential term in (A.1).
Note that the time dependence has been taken into account separately and
that two terms in the sum contribute to the indicated gradient.}

Now solve the AVTD equations (3.12) and (3.13) and their $\tau$
deriv\-a\-tives us\-ing
$\tilde Q_i$, $\tilde P_i$, $\tilde \pi _{Q_i}$, $\tilde \pi
_{P_i}$ for the constants $\zeta _i,\;\beta _i,\;\alpha _i,\;$and$\ \xi _i$ .
We find

{\parindent=0pt 1.  If $\tilde \pi _{P_i}\ne 0\ and\ \tilde \pi _{Q_i}\ne 0$
then
$$\zeta _i={{-1+\sqrt {1+\left( {{{\tilde \pi _{Q_i}e^{-\tilde P_i}} \over
{\tilde \pi _{P_i}}}} \right)^2}} \over {\left( {{{\tilde \pi _{Q_i}e^{-\tilde
P}} \over {\tilde \pi _{P_i}}}} \right)}},\eqno(A.3a)$$

$$\beta _i=\left( {{{1+\zeta _i^2} \over {\zeta _i^2-1}}} \right)\tilde \pi
_{P_i},\eqno(A.3b)$$

$$\alpha _i={{e^{\tilde P_i}} \over {1+\zeta _i^2}},\eqno(A.3c)$$

$$\xi _i=\tilde Q_i+\zeta _ie^{-\tilde P_i}.\eqno(A.3d)$$

2.  If $\tilde \pi _{P_i}\ne 0\ but\ \tilde \pi _{Q_i}=0$ then
$$\zeta _i=0,\;\beta _i=-\tilde \pi _{P_i},\;\alpha _i=e^{\tilde P_i},\;\xi
_i=\tilde Q_i.\eqno(A.4)$$

3.  If $\tilde \pi _{P_i}=0$ then for any $ \tilde \pi _{Q_i},$
$$\eqalign{&\zeta _i=1,\;\beta _i=-\tilde \pi _{Q_i}e^{-\tilde P_i},\cr
  &\alpha _i={\textstyle{1 \over 2}}e^{\tilde P_i},\;\xi _i=\tilde Q_i+e^{-
\tilde P_i}\quad.\cr}\eqno(A.5)$$

These values are then used to evolve the ${\tilde Q_i,\tilde P_i,\tilde \pi
_{Q_i},\tilde \pi _{P_i}}$ with the AVTD equations to $\tau ^j+\Delta
\tau $.  This is just the evolution by $H_1$ required by the algorithm. We
find
$${\tilde {\tilde \pi}} _{Q_i}=\tilde \pi _{Q_i},\eqno(A.6a)$$

$${\tilde {\tilde \pi}} _{P_i}=-{{\beta _i\left( {e^{-\beta _i(\tau _{j+1}-
\tau
_j)}-\zeta _i^2e^{\beta _i(\tau _{j+1}-\tau _j)}} \right)} \over {\left( {e^{-
\beta _i(\tau _{j+1}-\tau _j)}+\zeta _i^2e^{\beta _i(\tau _{j+1}-\tau _j)}}
\right)}},\eqno(A.6b)$$

$$\tilde {\tilde P_i}=\ln \left[ {\alpha _i\left( {e^{-\beta _i(\tau
_{j+1}-\tau
_j)}+\zeta _i^2e^{\beta _i(\tau _{j+1}-\tau _j)}} \right)}
\right],\eqno(A.6c)$$

$$\tilde {\tilde Q_i}=\xi _i-\left[ {{{\zeta _ie^{2\beta _i(\tau _{j+1}-\tau
_j)}} \over {\alpha _i\left( {1+\zeta _i^2e^{2\beta _i(\tau _{j+1}-\tau
_j)}} \right)}}} \right].\eqno(A.6d)$$

Finally, the ${\tilde {\tilde Q_i},\tilde {\tilde P_i},{\tilde {\tilde \pi}}
_{Q_i}
,{\tilde
{\tilde \pi }}_{P_i}}$ are evolved with $H_2$ for ${{\Delta \tau }
\mathord{\left/ {\vphantom {{\Delta \tau } 2}} \right. \kern-
\nulldelimiterspace} 2}$ to yield the original variables updated to the next
time-step:}

$$\eqalign{&Q_i(\tau ^{j+1})=\tilde {\tilde Q_i},\cr
  &P_i(\tau ^{j+1})=\tilde {\tilde P_i},\cr
  &\pi _{Q_i}(\tau ^{j+1})={\tilde {\tilde \pi}} _{Q_i}+{\textstyle{1
\over
2}}e^{-(\tau _{j+1}+\tau _j)}\left[ {e^{-(\tau _{j+1}-\tau _j)}-1}
\right]\left. {{{\partial V} \over {\partial Q_i}}} \right|_{{\tilde {\tilde
Q_k},\tilde {\tilde P_k}}},\cr
  &\pi _{P_i}(\tau ^{j+1})={\tilde {\tilde \pi}}_{P_i}+{\textstyle{1 \over
2}}e^{-(\tau _{j+1}+\tau _j)}\left[ {e^{-(\tau _{j+1}-\tau _j)}-1}
\right]\left. {{{\partial V} \over {\partial P_i}}} \right|_{{\tilde {\tilde
Q_k},\tilde {\tilde P_k}}}\quad.\cr}\eqno(A.7)$$

To achieve fourth order accuracy in time, repeat the entire
procedure three times with steps of $s\Delta \tau $, $(1-2s)\Delta \tau $,
and $s\Delta \tau $ respectively with $s=(2-2^{{1 \mathord{\left/
{\vphantom {1 3}} \right. \kern-\nulldelimiterspace} 3}})^{-1}$.  For
fourth order spatial accuracy, replace the potential term in (A.1) by the
appropriate version of (4.1).
\vfill
\eject
{\parindent=0pt REFERENCES

[1]  R.~Penrose, {\it Phys.~Rev.~Lett.~}{\bf 14}, 57 (1965);
S.W.~Hawking,
{\it Proc.~Roy.~Soc.~Lond.~A}{\bf 300}, 187 (1967); S.W.~Hawking
and R.~
Penrose, {\it Proc.~Roy.~Soc.~Lond.~A}{\bf  314}, 529 (1970); See also
S.W.~Hawking and G.F.R.~Ellis, {\bf The Large Scale Structure of
Space-Time} (Cambridge, Cambridge University, 1973); R.M.~Wald, {\bf
General Relativity }(Chicago, University of Chicago, 1984).

[2]  V.A. Belinskii, I.M. Khalatnikov, {\it Sov. Phys. JETP} {\bf 29},
911
(1969); V.A. Belinskii, I.M. Khalatnikov,  {\it Sov. Phys. JETP
}{\bf30},
1174 (1969); V.A. Belinskii, I.M. Khalatnikov,  {\it Sov. Phys. JETP
}{\bf
32}, 169 (1971); V.A. Belinskii, I.M. Khalatnikov, E.M. Lifshitz,  {\it
Adv.
Phys.} {\bf 19}, 525 (1970);   V.A. Belinskii, I.M. Khalatnikov, E.M.
Lifshitz,  {\it Adv. Phys.} {\bf 31}, 639 (1982);  V.A. Belinskii, E. M.
Lifshitz, I.M. Khalatnikov,  {\it Sov. Phys. Usp.} {\bf 13}, 745P765
(1971).

[3]  I.M.~Khalatnikov, V.L.~Pokrovski, in {\bf Magic without Magic},
edited by J.~Klauder, 441 (Freeman, San Francisco, l972).

[4]   C.W.~Misner,  {\it Phys.~Rev.~Lett.} {\bf 22}, 1071 (1969); C.W.~
Misner,  in {\bf Relativity}, edited by M.~Carmeli et al, 55 (Plenum, New
York, l970); C.W.~Misner,  in {\bf Magic without Magic}, edited by J.~
Klauder, 441 (Freeman, San Francisco, l972).

[5]  See M.~MacCallum in {\bf General Relativity, an Einstein
Centenary Survey}, edited by S.W.~Hawking and W.~Israel (Cambridge
University, Cambridge, 1979); R.T.~Jantzen, in {\bf Cosmology of the
Early Universe}, edited by R.~Ruffini and L.Z.~Fang (World Scientific,
Singapore, 1984), 233.

[6] See J.D.~Barrow and F.~Tipler, {\it Phys.~Rep.} {\bf 56}, 372 (1979).

[7]  P.~Halpern, {\it J.~Gen.~Rel.~Grav.} {\bf 19}, 73 (1987); R.M.Wald
and
A.~Higuchi, private communication; H.~Ishihara, {\it
Prog.~Theor.~Phys.}
{\bf 74}, 490 (1985).  Note that a rotational degree of freedom provides a
counterexample [see M.P.~Ryan, Jr.,  {\it Ann.~Phys.~(N.Y.)} {\bf 65},
506
(1971); M.P.~Ryan, Jr., L.C.~Shepley, {\bf Homogeneous Relativistic
Cosmologies} (Princeton University, Princeton, l975)].

[8] D.~Eardley, E.~Liang, R.~Sachs, {\it J.~Math.~Phys.} {\bf 13}, 99
(1972).

[9]  J.~Isenberg, V.~Moncrief,  {\it Ann.~Phys.~(N.Y.)} {\bf 199}, 84
(1990).

[10]  R.H.~Gowdy,  {\it Phys.~Rev.~Lett.} {\bf 27}, 826, erratum 1102
(1971);
{\it Ann.~Phys.~(N.Y.)} {\bf 83}, 203 (1974).

[11]  P.T.~Chrusciel, J.~Isenberg, V.~Moncrief, {\it
Class.~Quant.~Grav.}
{\bf 7}, 1671 (1990).

[12]  B.K.~Berger,  {\it Ann.~Phys.~(N.Y.)} {\bf 83}, 458 (l974).

[13]  B.~Grubi\v si\'c and V.~Moncrief, {\it Phys.~Rev.~D} {\bf 47},
2371
(1993).

[14]  V.~Moncrief,  {\it Ann.~Phys.~(N.Y.)} {\bf 167}, 118 (1986).

[15]  J.~Eells and J.~Sampson, {\it Amer.~J.~Math.} {\bf 86}, 109
(1964); C.~
Misner, {\it Phys.~Rev.~D} {\bf 18}, 4510 (1978).

[16]  V.~Moncrief,  {\it Ann.~Phys.~(N.Y.)} {\bf 132}, 87 (1981).

[17]  J.~A.~Fleck, J.~R.~Morris, and M.~D.~Feit, {\it  Appl.~Phys.} {\bf
10},
129 (1976); V.~Moncrief, {\it  Phys.~Rev.~D} {\bf 28}, 2485 (1983).

[18]  B.K.~Berger, D.~Garfinkle, and V.~Moncrief, unpublished.

[19] J.~Centrella, R.A.~Matzner,  {\it Astrophys.~J.} {\bf 230}, 311
(1979);
{\it Phys.~Rev.~D} {\bf 25}, 930 (1982); P.~Anninos, J.~Centrella, R.A.~
Matzner, {\it Phys.~Rev.~D} {\bf 39}, 2155 (1989); {\it Phys.~Rev.~D}
{\bf 43}, 1808 (1991); {\it  Phys.~Rev.~D} {\bf 43}, 1825 (1991).

[20]  See  {\bf Frontiers in Numerical Relativity}  edited by C.~R.~
Evans, L.S.
Finn, D.W.~Hobill, (Cambridge, Cambridge University, 1989).~

[21]  M.~Suzuki, {\it Phys.~Lett.} {\bf A146}, 319 (1990); {\it J.~Math.~
Phys.} {\bf 32}, 400 (1991).

[22]  M.P.~Ryan, Jr., ``Plane-Symmetric Cosmologies---An Overview,'' in
SILARG, pp.~96--121.

[23] W.H.~Press, B.P.~Flannery, S.A.~Teukolsky, W.T.~Vetterling, {\bf
Numerical Recipes in FORTRAN: the Art of Scientific Computing
2nd edition} (Cambridge University, Cambridge, 1992); C.A.~Hall and
T.A.~Porsching, {\bf Numerical Analysis of Partial Differential
Equations} (Englewood Cliffs, NJ, Prentice Hall, 1990); A.R.~Mitchell
and D.F.~Griffiths, {\bf The Finite Difference Method in Partial
Differential Equations} (New York, Wiley, 1980).

[24]  R.~Ove, Ph.~D.~Thesis, Yale University, 1985.

[25]  B.K.~Berger,  {\it Ann.~Phys.~(N.~Y.)} {\bf 156}, 155 (l984).

[26] E.~Kasner, {\it Am.~J.~Math.} {\bf 43}, 130 (1921).

[27] R.A.~Isaacson, {\it Phys.~Rev.} {\bf 166}, 1263 (1968);  1272
(1968).

[28]  D.R.~Brill and J.B.~Hartle, {\it  Phys.~Rev.~B } {\bf 135}, 271
(1964).

[29]  This was noticed numerically by David Garfinkle, private
communication.

[30]  One can show that the spacetimes discussed by V. Moncrief in {\it
Ann.~Phys.~(N.Y.)} {\bf 141}, 83 (1982), restricted to the Gowdy class,
have this propoerty.  For a more complete treatment of this question see P.
Mansfield, Ph.~D. Thesis, Yale University, 1989.

[31]  The exact solutions were computed using Bessel function
subroutines given in W.H. Press, B.P. Flannery, S.A. Teukolsky, W.T.
Vetterling, {\bf Numerical Recipes in FORTRAN: the Art of Scientific
Computing 2nd edition} (Cambridge University, Cambridge, 1992).

[32]  P.~ Chrusciel, private communication.

[33]  M.W.~Choptuik, {\it Phys.~Rev.~Lett.} {\bf 70}, 9 (1993).

[34] A.M.~Abrahams and C.R.~Evans,  {\it Phys.~Rev.~Lett.} {\bf 70},
2980 (1993).

[35]  C.M.~Swift, ``Numerical Studies of the Gowdy T$^3$ Spacetime,''
Masters Thesis, Oakland University, 1992; B.K.~Berger, D.~Garfinkle,
C.M.~Swift, unpublished.

[36]  See the discussion of adaptive gridding in [33] and references
therein.

[37] R.~Arnowitt, S.~Deser, C.W Misner, in {\bf Gravitation:  An
Introduction to Current Research}, edited by L.~Witten (Wiley, New
York, l962), 227.

[38]  V.~Moncrief, unpublished.}
\vfill
\eject
{\parindent=0pt
\parskip=18pt FIGURE CAPTIONS

Fig. 1.  The exact solution for the pseudo-unpolarized test case.   Initial
data are generated from  Eq. (3.20) applied to (4.2) (and its $\tau$
derivative) evaluated at $\tau = 0$.  The axis scales for $\theta$ and $\tau$
are $[0, 2\pi]$ and $[0, 23]$ respectively.  The vertical axis indicates the
values of (a)  $P(\theta,\tau)$ and (b)  $Q(\theta,\tau)$.

Fig. 2.  Errors in the $\theta$-$\tau$ plane for the 2nd order accurate
numerical scheme.  The ranges of $\theta$ and $\tau$ are the same as in
Fig.~1.  (a) The vertical scale measures $P_{numerical} - P_{exact}$
(where
$P_{exact}$ is shown in Fig.~1(a)) and $P_{numerical} $ is the
computation.  (b)
The same as (a) but for $Q$.

Fig. 3.  Errors in the $\theta$-$\tau$ plane for the 4th order accurate
numerical scheme.  All axes and variables are the same as in Fig.~2 except
that the 4th order accurate numerical scheme has been used.

Fig. 4.  $P(\theta,\tau)$ vs.~$\theta$ at selected values of $\tau$.  Results
of the simulation for initial data from (5.1) in the 4th order accurate
numerical scheme are shown.  The computation was performed with 800
spatial grid points.  The display consists of values at 100 of these grid
points averaged over nearest and next nearest neighbors in the full array.
(The selection of spatial grid points accounts for any asymmetry about
$\theta=\pi$.)  Six graphs at selected $\tau$ values are stacked.  In all
cases, the horizontal axis is $0 \le\theta\le 2\pi$.  The numerical scales on
either the left or right axis denote the amplitude of $P$.  The simulation
represents the range $0 \le \tau \le 18.85$.

Fig. 5.  $QJ(\theta,\tau)$ vs. $\theta$ at selected values of $\tau$.  The
same as Fig.~4 but for $Q$.  For $\tau \ge 1.14$, the vertical scale is $[0,
500]$.

Fig. 6.  $v(\theta,\tau)$ vs. $\theta$ at selected values of $\tau$.  The same
as Fig.~4 but for $v$ as defined by (3.21).

Fig. 7.  $P(\theta,\tau)$,  $Q(\theta,\tau)$,  and $v(\theta,\tau)$  vs.
$\tau$ at selected values of $\theta$.  In all cases, the horizontal axis is $0
\le \tau \le 18$.  On the graph for $Q$, the values of the solid lines
correspond to the left axis and the dashed lines to the right axis.

Fig. 8.  $P$ and $Q$ in the $\theta$-$\tau$ plane.  The complete results of
the simulation in Figs.~4--7 are shown in the $\theta$-$\tau$ plane.  The
axis scales for $\theta$ and $\tau$ are $[0, 2\pi]$ and $[0, 6\pi]$
respectively.  The vertical axes denote the values of (a) $P$ and (b) $Q$
respectively.

Fig. 9.  The values of $v$ in the $\theta$-$\tau$ plane.  This figure is the
same as Fig.~8 but for the parameter $v$.   Note that the viewing angle
has
changed for greater clarity.  (a)  All values of $v$ .  (b)  The same data
with the vertical scale chosen to display only $v \le 1$.

Fig. 10.  A ``non-generic'' point in $\theta$.  $Q$ and $v$  vs.~$\theta$ at
$\tau = 6\pi$ are shown near a point with $v > 1$ for the initial data of
Figs.~4--9 for a simulation with 6400 spatial grid points with no spatial
averaging of the results.  The graphs are produced using all the spatial
resolution available in the simulation.  The left axis corresponds to $v$
and the right to $Q$.  Note that $v > 1$ only where $Q,_\theta \approx 0$
and that the ranges displayed for $Q$ and $\theta$ are small.  The
horizontal dashed line denotes $v = 1$.

Fig. 11.  Approaching the AVTD limit.  Graphs of $v$ (solid line) and
$P/\tau$ (dashed line) vs.~$\theta$ are overlaid.  The data are taken from
the same simulation as Figs.~4--9 for 800 spatial grid points but with
no spatial averaging.  The vertical scale for both $P/\tau$ and $v$ ranges
from 0 to 1.2 while the horizontal axis is $2.6 \le \theta \le 3.2$.  (a)
$\tau = 18.86$; (b) $\tau = 43.96$.

Fig. 12.  Growth of small scale spatial structure.  The graph shows the data
for $P(\theta,\tau)$ from Fig.~8(a) for the range $0 \le \tau \le 4.2$.

Fig. 13.  Spatial resolution dependence of generic spiky features.  A
typical spiky feature in $\pi_P$ vs.~$\theta$ at $\tau = 2\pi$  is shown at
resolutions of 400, 800, 1600, and 6400 spatial grid points.

Fig. 14.  Evolution of a spiky feature.  The dashed line indicates the same
plot as Fig.~13 at $\tau = 2\pi$ for 6400 spatial grid points on the
expanded scale of this figure.  The same feature at 6400 spatial grid points
is shown at $\tau = 4\pi$ (solid line) and $\tau = 6\pi$ (dashed line with
circles).  For comparison, the feature computed with 1600 spatial grid
points is shown at $\tau = 4\pi$ (dashed line with squares).}

\end